\documentclass[11pt,english,aps,preprint,nofootinbib]{revtex4}
\usepackage[T1]{fontenc}
\usepackage[latin1]{inputenc}
\usepackage{amsmath}
\usepackage{graphicx}
\usepackage{amssymb}

\makeatletter

\usepackage{graphicx}

\newcommand{\ssst}[1]{\scriptscriptstyle{#1}}

\newcommand{\bra}[1]{\langle #1 \vert}
\newcommand{\ket}[1]{\vert #1 \rangle}

\newcommand{\mN}{m_{\ssst{N}}}

\newcommand{\gpiNN}{g_{\pi\ssst{NN}}}
\newcommand{\grhoNN}{g_{\rho\ssst{NN}}}
\newcommand{\gomegaNN}{g_{\omega\ssst{NN}}}

\usepackage{bm}
\usepackage{color}

\allowdisplaybreaks

\def\be{\begin{equation}}
\def\ee{\end{equation}}
\def\bea{\begin{eqnarray}}
\def\eea{\end{eqnarray}}

\def\vsig{\bm{\sigma}}
\def\vtau{\bm{\tau}}
\def\hpiNN{h_\pi^1}

\def\mbfa{\bm{a}}

\def\mbfI{\bm{I}}
\def\mbfj{\bm{j}}

\def\mbfl{\bm{l}}
\def\mbfp{\bm{p}}
\def\mbfP{\bm{P}}
\def\mbfr{\bm{r}}
\def\mbfR{\bm{R}}
\def\mbfx{\bm{x}}

\def\S12{S_{12}(\hat{\mbfr})}
\def\vto{\tilde{v}_{3p1}(r)}
\def\voo{\tilde{v}_{1p1}(r)}

\usepackage{babel}
\makeatother
\begin{document}

\title{Deuteron Anapole Moment with Heavy Mesons}

\author{C.-P. Liu$^{1}$}

\email{cpliu@triumf.ca}

\author{C. H. Hyun$^{2}$}

\email{hch@meson.skku.ac.kr}

\author{B. Desplanques$^{3}$}

\email{desplanq@lpsc.in2p3.fr}

\affiliation{$^{1}$TRIUMF, 4004 Wesbrook Mall, Vancouver, British Columbia, Canada 
V6T 2A3\\
$^{2}$Institute of Basic Science, Sungkyunkwan University, Suwon
440-746, Korea\\
$^{3}$Laboratoire de Physique Subatomique et de Cosmologie\\
(UMR CNRS/IN2P3-UJF-INPG), F-38026 Grenoble Cedex, France}

\date{\today}

\begin{abstract}
Parity-nonconserving two-body currents due to vector meson exchange
are considered with the aim to determine the related contributions
to the anapole moment. A particular attention is given to the requirement
of current conservation which is essential for a reliable estimate
of this quantity. An application is made for the deuteron case.
\end{abstract}
\maketitle

\section{Introduction \label{sec:intro}}

First introduced by Zel'dovich \cite{Zeld57}, the anapole moment
(AM) of a quantum system is a quantity that involves both the electromagnetic
interaction and parity nonconservation (PNC). Ignored for a long time,
it is not before studies by Flambaum and Khriplovich \cite{FlKr80}
that the concept acquires practical interest. In their work, these
authors especially emphasize that the AM would grow with the size
of the nucleus, making heavy nuclei natural candidates for an observation.
Implying the hyperfine structure, a first measurement was performed
a few years ago in the $^{133}$Cs nucleus \cite{Colorado97}.

The deuteron AM has also received some attention recently 
\cite{SaSp98,SaSp01,Sa01,KhKo00,HyDe03}.
Though its interest is largely academic (an experiment is not feasible
in a near future), it offers the advantage of a laboratory where methods
and ingredients relative to an estimate can be studied in details.
These studies have been concerned with the pion-exchange component
of the PNC nucleon-nucleon (NN) force. Calculations were based on
assuming an effective-field-theory description \cite{SaSp98,SaSp01}, zero-range
NN strong forces \cite{KhKo00}, or more realistic NN strong forces
\cite{HyDe03}. The use of an alternative field-theory description
was also proposed \cite{Sa01}. The next step concerns the extension
of these results to include the component of the PNC force due to
vector-meson exchange that could contribute as much as, if not more
than the pion exchange. 

Determining the AM supposes to calculate the effective current that
couples to the photon. As it involves parity nonconservation, the
current has necessarily an axial character, making the requirement
of current conservation non-trivial. Individual contributions are proportional
to the weak coupling, $G_F$, while fulfilling the above property implies that 
the effective current contains the factor $G_F\,q^2$, which vanishes in the limit 
of a zero momentum transfer. Getting this result demands
particular care with contributions ensuring gauge invariance. When
dealing with vector mesons, this task becomes essential. In particular,
it has to be done consistently with the PNC interaction model that
is employed in calculations. Some contributions were given in Ref.
\cite{HLRM02}. In the present work, we intend to complete this work
with the double aim to satisfy gauge invariance and consistency with
the PNC interaction model; DDH potential, given by Desplanques, Donoghue,
and Holstein, is our case \cite{DDH80}. Some estimates of the vector-meson-exchange
contributions to the anapole moment will be presented in the deuteron
case. 

The plan of the paper is as follows. In Section \ref{sec:MECs}, we
first present the various ingredients pertinent to the interaction:
parity-conserving (PC), parity-nonconserving and electromagnetic (EM)
ones. We subsequently provide the expressions for the PNC two-body
currents at the lowest $1/\mN$ order and show how they allow one
to fulfill current conservation. Section \ref{sec:d AM} is devoted
to applications involving the deuteron. This includes the deuteron
description, especially the determination of the PNC components; the
expression of the anapole matrix elements from both the one- and two-body
currents; and a numerical estimate in terms of the PNC meson-nucleon
coupling constants. A discussion of the results is given in Section
\ref{sec:dis}. This is completed by Appendix \ref{sec:MECs in x}
that contains expressions of the two-body currents in configuration
space.

\section{PNC $NN$ Interaction, Currents, and Current Conservation \label{sec:MECs}}

The anapole moment is a special electromagnetic property
of a system in which parity conservation is violated; therefore, our
first step in calculation is formulating the EM current operators.
Throughout this work, we assume for a nuclear system the validity
of non-relativistic (NR) limit and keep only terms of leading order.

The one-body term $(\rho^{(1)},\bm j^{(1)})$, that is, contributions
from each individual nucleon (for deuteron, there are two), include
the charge density:\begin{equation}
\rho^{(1)}=e\sum_{i=1}^{2}\frac{1+\tau_{i}^{z}}{2}(2\pi)^{3}\delta^{(3)}
(\bm k+\bm p'{}_{i}-\bm p{}_{i})\,,\end{equation}
and the 3-current densities from spin and motion of nucleons:

\begin{eqnarray}
\bm j_{spin} & = & e\sum_{i=1}^{2}\frac{\mu_{\ssst{N}}^{i}}{2\mN}i\bm\sigma_{i}\times
(\bm p'_{i}-\bm p_{i})(2\pi)^{3}\delta^{(3)}(\bm k+\bm p'{}_{i}-\bm p{}_{i})\,,\\
\bm j_{conv} & = & e\sum_{i=1}^{2}\frac{1+\tau_{i}^{z}}{4\mN}(\bm p'_{i}+\bm p_{i})(2\pi)^{3}\delta^{(3)}(\bm k+\bm p'{}_{i}-\bm p{}_{i})\,,\end{eqnarray}
 in momentum space, where $\mu_{\ssst{N}}^{i}$ is defined as \[
\mu_{\ssst{N}}^{i}\equiv\frac{1}{2}\left(\mu_{\ssst{S}}+\tau_{i}^{z}
\mu_{\ssst{V}}\right)\]
 with $\mu_{\ssst{S}}=0.88$ and $\mu_{\ssst{V}}=4.71$; and $\bm k$,
$\bm p'_{i}$, and $\bm p{}_{i}$ denote the 3-momentum of outgoing
photon, outgoing $i$th nucleon, and incoming $i$th nucleon, respectively.

As the canonical picture of nucleon-nucleon ($NN$) interaction is
realized through the exchange of mesons, the nuclear EM currents
should also contain these exchange effects, which are two-body in
character. Since we are still lacking of a fundamental understanding
about the $NN$ interaction, the exact exchange currents (ECs) are
also unknown. However, in order to reduce theoretical uncertainties
in the studies of nuclear EM processes, it is important, given a chosen
model for the $NN$ interaction, to construct these two-body currents
which are constrained by both current conservation and phenomenology.

For the PC $NN$ interaction, many high-quality
model potentials exist, and our choice for calculation in this work
is the Argonne $v{}_{18}$ potential (A$v{}_{18}$) \cite{AV18}.
Although it gives good fits to the scattering data and deuteron properties,
it is not straightforward to construct the corresponding ECs because the
connection with the meson exchange picture is not clear for some parts
in this potential. One traditional way to construct the ECs is implementing
the NR minimal coupling (MC) to the potential, i.e.,\[
\bm p\rightarrow\bm p-\frac{e}{2}(1+\tau^{z})\bm A\,;\quad H\rightarrow H+\frac{e}{2}(1+\tau^{z})A^{0}\,,\]
 then identifying the EM currents from the interaction Hamiltonian
density, $ej_{\mu}A^{\mu}$. However, this procedure only constrains
the longitudinal components, while giving no information about the
transverse components which are conserved by themselves. 
Some uncertainty about these last terms comes from potentials involving 
quadratic velocity-dependent components as discussed in Ref. \cite{Risk89}. 
Moreover, the derivation of exchange currents for a model like A$v_{18}$, 
employed in Ref. \cite{Scetal03} for instance, requires further elaboration. This 
model contains a Gaussian type component while the above prescription is usually 
applied to Yukawa potentials.
Therefore, we leave this as an open question
for future work and follow the treatment of Ref. \cite{HyDe03}
to examine: (1) to what degree current conservation is broken
by the omission of PC ECs, and (2) how much the inclusion of PC ECs
due to the one pion exchange, which gives the long-range part in
A$v{}_{18}$, could restore the conservation. By this exercise, one
can get some qualitative handle on this problem. The detail is be
to discussed in Section \ref{sec:dis}. 

For the PNC $NN$ interaction, our choice is
the potential suggested by Desplanques, Donoghue, and Holstein (DDH)
\cite{DDH80}. Since this potential, based on the one-boson exchange
scheme involving $\pi$, $\rho$, and $\omega$ mesons, has a close
tie with the exchange picture, a more field-theoretical formalism,
the so-called $S$-matrix approach \cite{ChRh70,Chem79,Town87,HLRM02},
is used to construct all the corresponding ECs. As we will show later,
some transverse components arise naturally in this derivation. For
clarity, we divide the following discussion into three parts: first,
the model Lagrangian, consistent with the DDH scheme, is constructed;
second, the PNC ECs are derived; and finally, we show how these 
exchange currents allow one to fulfill current conservation, given the DDH 
potential.

\subsection{Model Lagrangian}

The total Lagrangian density we consider is expressed as

\begin{equation}
\mathcal{L}=\mathcal{L}_{0}+\mathcal{L}_{PC}+\mathcal{L}_{PNC}+\mathcal{L}_{EM}
+\delta\mathcal{L}\,,\end{equation}
where $\delta\mathcal{L}$ contains all the terms not relevant for
this discussion. The free Lagrangian densities for nucleon, $N$;
pion, rho, and omega mesons, $\bm\pi$, $\bm\rho^{\mu}$, and $\omega^{\mu}$,
are\begin{eqnarray}
\mathcal{L}_{0} & = & \bar{N}'(i\partial\!\!\!/-\mN)N+\frac{1}{2}(\partial_{\mu}\bm\pi)\cdot
(\partial^{\mu}\bm\pi)-\frac{1}{2}m_{\pi}^{2}\bm\pi^{2}-\frac{1}{4}\bm F_{\mu\nu}^{(\rho)}\cdot\bm F^{(\rho)\mu\nu}+\frac{1}{2}m_{\rho}^{2}\bm\rho_{\mu}\cdot\bm\rho^{\mu}-
\frac{1}{2\xi}(\partial_{\mu}\bm\rho^{\mu})\cdot(\partial_{\nu}
\bm\rho^{\nu})\nonumber \\
 &  & -\frac{1}{4}F_{\mu\nu}^{(\omega)}F^{(\omega)\mu\nu}+\frac{1}{2}
m_{\omega}^{2}\omega_{\mu}\omega^{\mu}-\frac{1}{2\xi}(\partial_{\mu}\omega^{\mu})
(\partial_{\nu}\omega^{\nu})\,,\end{eqnarray}
where $\bm F_{\mu\nu}^{(\rho)}$ and $F_{\mu\nu}^{(\omega)}$ are
the field tensors of the vector mesons, and we keep the $R_{\xi}$
gauge-fixing terms of vector mesons explicit for the moment. The PC
and PNC meson-nucleon interaction Lagrangian densities are

\begin{eqnarray}
\mathcal{L}_{PC} & = & i\gpiNN\bar{N}'\gamma_{5}\bm\tau\cdot\bm\pi N-\grhoNN\bar{N}'(\gamma_{\mu}-i\frac{\chi_{\ssst{V}}}{2\mN}
\sigma_{\mu\nu}q^{\nu})\bm\tau\cdot\bm\rho^{\mu}N\nonumber \\
 &  & -\gomegaNN\bar{N}'(\gamma_{\mu}-i\frac{\chi_{\ssst{S}}}{2\mN}\sigma_{\mu\nu}
q^{\nu})\omega^{\mu}N\,,\\
\mathcal{L}_{PNC} & = & -\frac{h_{\pi}^{1}}{\sqrt{2}}\bar{N}'(\bm\tau\times\bm\pi)^{z}N+
\bar{N}'[h_{\rho}^{0}\bm\tau\cdot\bm\rho^{\mu}+h_{\rho}^{1}\rho^{z\mu}+
\frac{h_{\rho}^{2}}{2\sqrt{6}}(3\tau^{z}\rho^{z\mu}-\bm\tau\cdot\bm\rho^{\mu})]
\gamma_{\mu}\gamma_{5}N\nonumber \\
 &  & +\bar{N}'(h_{\omega}^{0}\omega^{\mu}+h_{\omega}^{1}\tau^{z}\omega^{\mu})
\gamma_{\mu}\gamma_{5}N\,,\end{eqnarray}
where $q^{\mu}$ is the 4-momentum carried by the outgoing boson;
the strong couplings, $g_{\ssst{XNN}}$, as well as the weak couplings,
$h_{\ssst{X}}^{(i)}$, are of DDH's definition (except their PNC pion
coupling, $f_{\pi}$, is renamed as $h_{\pi}^{1}$ here); and the
anomalous strong isoscalar and isovector magnetic moments of nucleon,
$\chi_{\ssst{S}}$ and $\chi_{\ssst{V}}$, are assumed to be the same
as the EM values, -0.12 and 3.70, by vector meson dominance.

Now we apply the covariant MC:\[
p_{\mu}\rightarrow p_{\mu}-\frac{e}{2}(1+\tau^{z})A_{\mu}\,,\]
 to the above Lagrangian densities to obtain the EM interactions.
For our purpose, only terms of first-order in $e$ are included in
$\mathcal{L}_{EM}$. From $\mathcal{L}_{0}$, we get

\begin{eqnarray}
\mathcal{L}_{EM}^{(\ssst{NN}\gamma)} & = & -e\bar{N'}[(F_{1}^{\ssst{(S)}}(Q^{2})\frac{1}{2}+F_{1}^{\ssst{(V)}}(Q^{2})
\frac{\tau^{z}}{2})\gamma_{\mu}-i\frac{1}{2\mN}(F_{2}^{\ssst{(S)}}(Q^{2})
\frac{1}{2}+F_{2}^{\ssst{V}}(Q^{2})\frac{\tau^{z}}{2})\sigma_{\mu\nu}q^{\nu}]N
A^{\mu}\,,\\
\mathcal{L}_{EM}^{(\pi\pi\gamma)} & = & -e(\bm\pi\times\partial_{\mu}\bm\pi)^{z}A^{\mu}\,,\\
\mathcal{L}_{EM}^{(\rho\rho\gamma)} & = & -e(\bm\rho^{\nu}\times\bm F_{\nu\mu}^{(\rho)})^{z}A^{\mu}-\frac{1}{\xi}e(\bm\rho_{\mu}\times\partial_{\nu}
\bm\rho^{\nu})^{z}A^{\mu}\,.\end{eqnarray}
Note that in order to account for the nucleon structure, nucleon EM
form factors, $F_{1,2}^{\ssst{(S,V)}}$ (superscript "$S$" for isoscalar
and "$V$" for isovector; subscript "$1$" for Dirac and "$2$" for Pauli)
have to be added. At $Q^{2}=-q^{2}=0$, $F_{1}^{\ssst{(S)}}(0)=F_{1}^{\ssst{(V)}}(0)=1$,
$F_{2}^{\ssst{(S)}}(0)=-0.12$, and $F_{2}^{\ssst{(V)}}(0)=3.70$.
In principle one should also take into account the meson structures;
however, they are still poorly constrained so we simply assume them
elementary. 

Due to the momentum-dependent coupling of $\rho$ meson to the nucleon
anomalous magnetic moment, a Kroll-Ruderman type contact interaction
\cite{KrRu54}, reading as\begin{equation}
\mathcal{L}_{EM}^{(\ssst{NN}\rho\gamma)}=-e\frac{\grhoNN\chi_{\ssst{V}}}{2\mN}
\bar{N'}\sigma_{\mu\nu}(\bm\tau\times\bm\rho^{\nu})^{z}NA^{\mu}\,,\end{equation}
also arises. This will lead to a seagull current which is important
for current conservation, but was ignored in Ref. \cite{HLRM02}.

It is worthwhile to point out that the EM interactions obtained above
depend on the model Lagrangians we start out. For example, compared
with the result of QHD II \cite{SeWa86}, a gauge theory approach
which first models pion from a global $SU(2)$ symmetry group and
then adds $\rho$ meson and photon by a local $SU(2)\otimes U(1)$ symmetry
group (both steps involve Higgs mechanism to generate meson masses),
one observes a larger $\rho\rho\gamma$ interaction in QHD II by the amount of
\[\Delta\mathcal{L}_{EM}^{(\rho\rho\gamma)}=
\frac{1}{2}e(\bm\rho_{\mu}
\times\bm\rho_{\nu})^{z}F^{(\gamma)\mu\nu}\,,\]
which modifies the $\rho\rho\gamma$ vertex in an interesting way as we
are going to explain. 

\begin{figure}
\includegraphics[%
  scale=0.6]{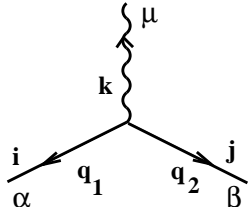}

\caption{The vertex factor for vector-meson-photon coupling, where $\alpha$,
$\beta$, and $\gamma$ are the Lorentz indices; $i$ and $j$ are
isospin indices. \label{fig:vertex}}
\end{figure}

By fixing the gauge parameter $\xi=1$, i.e., 't Hooft-Feynman gauge,
we obtained the same $\mathcal{L}_{EM}^{(\rho\rho\gamma)}$ as in
Ref. \cite{Hyetal80}, and this gives a vertex factor (see Fig. \ref{fig:vertex}):\begin{equation}
\epsilon_{3ij}[(q_{1}-q_{2})^{\mu}g^{\alpha\beta}+k^{\alpha}g^{\beta\mu}-
k^{\beta}g^{\mu\alpha}]\,,\label{eq:rho MC}\end{equation}
with $k+q_{1}+q_{2}=0$. The latter two terms are combined to give
amplitudes conserved by themselves when dotted by $k_{\mu}$, which
means they are purely transverse (actually, they correspond to magnetic
dipole couplings which explain the transversality). By adding $\Delta\mathcal{L}_{EM}^{(\rho\rho\gamma)}$,
it simply doubles the self-conserved terms, so that we have\begin{equation}
\epsilon_{3ij}[(q_{1}-q_{2})^{\mu}g^{\alpha\beta}+2k^{\alpha}g^{\beta\mu}-
2k^{\beta}g^{\mu\alpha}]\,,\label{eq:rho chiral}\end{equation}
for the vertex. As $\rho$ meson is a spin-one particle, it has charge
($c$), magnetic dipole ($\mu$), and charge quadrupole ($Q$) couplings
to the EM field. When assuming it is an elementary particle, $c=e$,
$\mu=e/m_{\rho}$, and $Q=-e/m_{\rho}^{2}$, the vertex factor appears
to be Eq. (\ref{eq:rho chiral}) \cite{Aretal80,BrHi92}, so the MC
result under-predicts the $\rho$ meson magnetic moment by two. This factor
of two difference in self-conserved terms between MC and chiral Lagrangian
approaches has been pointed out in Refs. \cite{AdTr84,Town87}. It is also 
obtained from a quark model calculation  of the charged $\rho$ meson 
magnetic moment (in the limit where the $\rho$ mass 
is taken as twice the constituent quark one). Although the factor two 
originates from different models, in order to have a closer
contact with phenomenology, we make this particular modification to
the $\rho\rho\gamma$ vertex.

The modification mentioned above is just an example of model-dependences
in constructing ECs. Since these purely transverse terms, often called
non-Born (NB) terms, could not be constrained by current conservation,
it is not easy to set up criteria \emph{a priori} to judge which should
be included or not, unless compared with experiments %
\footnote{There are a lot of discussions in literature addressing this issue.
We refer readers to more extensive reviews such as Refs. \cite{Chem79,Town87,Math89,Risk89}
(and references therein) for details.%
}. For our calculation, we choose to include $\rho\pi\gamma$ and $\omega\pi\gamma$
interactions \begin{eqnarray}
\mathcal{L}_{EM}^{(\rho\pi\gamma)} & = & e\frac{g_{\rho\pi\gamma}}{2\, m_\rho}\epsilon_{\alpha\beta\gamma\delta}F^{(\gamma)
\alpha\beta}(\bm\rho^{\gamma}\cdot\partial^{\delta}\bm\pi)\,,\\
\mathcal{L}_{EM}^{(\omega\pi\gamma)} & = & e\frac{g_{\omega\pi\gamma}}{2\, m_\omega}\epsilon_{\alpha\beta\gamma\delta}F^{(\gamma)
\alpha\beta}(\omega^{\gamma}\partial^{\delta}\pi^{z})\,,\end{eqnarray}
where the total anti-symmetric tensor is defined as $\epsilon_{0123}=-1$
\cite{Adle66,Chem79}, because these coupling constants could be determined
from the decay data (except for signs); and we ignore all the nucleon
isobaric excitations for they are not the main theoretical emphasis of this work.
The present work could be easily extended to them if necesssary.

To sum up, the total EM Lagrangian density we consider is\begin{equation}
\mathcal{L}_{EM}=L_{EM}^{(\ssst{NN}\gamma)}+L_{EM}^{(\pi\pi\gamma)}+\left\{ L_{EM}^{(\rho\rho\gamma)}+\frac{1}{2}e(\bm\rho_{\mu}\times\bm\rho_{\nu})^{z}
F^{(\gamma)\mu\nu}\right\} +\mathcal{L}_{EM}^{(\rho\pi\gamma)}+\mathcal{L}_{EM}^{(\omega\pi\gamma)}
\,.\end{equation}

\subsection{PNC Meson Exchange Currents}

\begin{figure}
\includegraphics[%
  scale=0.4]{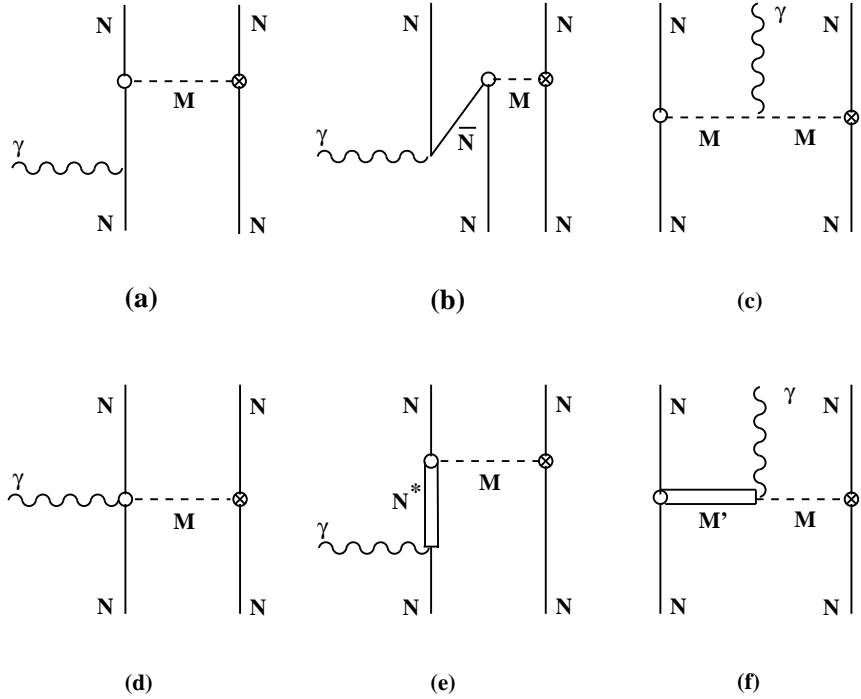}

\caption{Classification of meson exchange currents: (a) norm-recoil, (b) pair,
(c) mesonic, (d) seagull, (e) isobaric, and (f) non-Born mesonic,
where N and N{*} denote nucleon and nucleon excited state; M and M'
denote mesons. \label{fig:MECs}}
\end{figure}

Diagrammatically, the ECs could be classified, according to Fig. \ref{fig:MECs},
as: (a) norm-recoil, (b) pair, (c) mesonic, (d) seagull, (e) isobaric,
and (f) NB mesonic types. The division of norm-recoil and pair terms
simply comes from the separation of positive- and negative-energy
components in the covariant nucleon propagator. Sometimes confusion
arises when it comes to pair and seagull diagram. If the PC $\pi NN$
coupling is formulated as pseudo-vector, the seagull term is $O(1/\mN)$,
while the pair term is higher-order in $1/\mN$. On the other hand,
if the pseudo-scalar coupling is adopted as we do here, there is no
seagull term, however, at the leading order, the pair term looks exactly
as the seagull term in the pseudo-vector scheme. Therefore, as far
as the NR approximation is valid, these two formalisms are equivalent
\cite{Math89,Risk89}. For the case of $\rho$ meson, both pair and
seagull diagrams have $O(1/\mN)$ contributions. As for the NB contributions
from (e) and (f), we only consider the latter as explained above.

Before one applies Feynman rules to evaluate these diagrams and extract
the corresponding ECs, the gauge parameter has to be fixed. Though
the physical results should be gauge-independent, a proper choice
may greatly simplify the calculation. Here, we adopt the 't Hooft-Feynman
gauge, $\xi=1$, for the following reasons. First, the propagator
is simpler\begin{eqnarray*}
\langle A_{\mu}A_{\nu}\rangle & = & \left(g_{\mu\nu}-(1-\xi)\frac{q_{\mu}q_{\nu}}{q^{2}-\xi m^{2}+i\epsilon}\right)\frac{-i}{q^{2}-m^{2}+i\epsilon}\\
 & = & \frac{-ig_{\mu\nu}}{q^{2}-m^{2}+i\epsilon}\quad(\textrm{for }\xi=1).\end{eqnarray*}
Second, the PNC potential, constructed from NR reduction of the one-boson
exchange diagrams, corresponds to the form given by DDH. The last
and most important of all, in combination with the previous statement,
the contribution from the norm-recoil diagram represents how the one-body
EM matrix element is modified by the presence of this NR potential
\cite{Town87}. Therefore, it should not be double-counted if one
has already taken care of this by using the perturbed wave function,
the route we will follow. 

In momentum space, to the order of $1/\mN$, we have the following
results for the pair and $\rho$-seagull (KR) 3-currents

\begin{eqnarray}
\bm j_{pair}^{\pi} & = & \frac{-e\gpiNN h_{\pi}^{1}}{2\sqrt{2}\mN}(\bm\tau_{1}\cdot\bm\tau_{2}-\tau_{1}^{z}
\tau_{2}^{z})[\bm\sigma_{1}]\frac{(2\pi)^{3}\delta^{(3)}(\cdots)}{\bm q_{2}^{2}+m_{\pi}^{2}}+(1\leftrightarrow2)\,,\label{eq:PNC MEC first}\\
\bm j_{pair+\ssst{KR}}^{\rho} & = & \frac{-e\grhoNN}{2\mN}\left\{ \left(h_{\rho}^{0}(\bm\tau_{1}\cdot\bm\tau_{2}+\tau_{2}^{z})+
\frac{h_{\rho}^{2}}{2\sqrt{6}}(3\tau_{1}^{z}\tau_{2}^{z}-
\bm\tau_{1}\cdot\bm\tau_{2}+2\tau_{2}^{z})\right)[\bm\sigma{}_{1}-
\bm\sigma_{2}]\right.\nonumber \\
 &  & \left.+h_{\rho}^{1}(1+\tau_{1}^{z})[\tau_{2}^{z}\bm\sigma_{1}-
\tau_{1}^{z}\bm\sigma_{2}]+(1+\chi_{\ssst{V}})\left(h_{\rho}^{0}-
\frac{h_{\rho}^{2}}{2\sqrt{6}}\right)(\bm\tau_{1}\times\bm\tau_{2})^{z}
[\bm\sigma_{1}\times\bm\sigma_{2}]\right\} \nonumber \\
 &  & \times\frac{(2\pi)^{3}\delta^{(3)}(\cdots)}{\bm q_{2}^{2}+m_{\rho}^{2}}+(1\leftrightarrow2)\,,\\
\bm j_{pair}^{\omega} & = & \frac{-e\gomegaNN}{2M_{N}}(1+\tau_{1}^{z})\big(h_{\omega}^{0}[\bm\sigma_{1}-
\bm\sigma_{2}]+h_{\omega}^{1}[\tau_{1}^{z}\bm\sigma_{1}-\tau_{2}^{z}
\bm\sigma_{2}]\big)\frac{(2\pi)^{3}\delta^{(3)}(\cdots)}{\bm q_{2}^{2}+m_{\omega}^{2}}+(1\leftrightarrow2)\,,\end{eqnarray}
where the $\rho$-seagull 3-current corresponds to the term involving
$\chi_{\ssst{V}}$; $\bm q_{1,2}=\bm p'_{1,2}-\bm p_{1,2}$; and the
$\delta$ function imposes the total 3-momentum conservation, $\bm k+\bm q_{1}+\bm q_{2}=0$.
All the pair charges are of higher order in $1/\mN$ compared with
the nucleon charge, which is $O(1)$, so they are neglected. 

For the mesonic 3-currents, we obtain\begin{eqnarray}
\bm j_{mesonic}^{\pi} & = & \frac{-e\gpiNN h_{\pi}^{1}}{2\sqrt{2}\mN}(\bm\tau_{1}\cdot\bm\tau_{2}-\tau_{1}^{z}\tau_{2}^{z})
[\bm q_{2}-\bm q_{1}]\bm\sigma_{1}\cdot\bm q_{1}\frac{(2\pi)^{3}\delta^{(3)}(\cdots)}{(\bm q_{1}^{2}+m_{\pi}^{2})(\bm q_{2}^{2}+m_{\pi}^{2})}+(1\leftrightarrow2)\,,\\
\bm j_{mesonic}^{\rho} & = & \frac{-e\grhoNN}{2\mN}\left(h_{\rho}^{0}-\frac{h_{\rho}^{2}}{2\sqrt{6}}\right)
i(\bm\tau_{1}\times\bm\tau_{2})^{z}\Big\{[\bm q_{1}-\bm q_{2}]\bm\sigma_{2}\cdot\big((\bm p'_{2}+\bm p_{2})-(\bm p'_{1}+\bm p_{1})-i(1+\chi_{\ssst{V}})\bm\sigma_{1}\times\bm q_{1}\big)\nonumber \\
 &  & +2[(\bm p'_{1}+\bm p_{1})+i(1+\chi_{\ssst{V}})\bm\sigma_{1}\times\bm q_{1}]\bm\sigma_{2}\cdot\bm k+2[\bm\sigma_{2}]\Big(2\mN k_{0}-\bm k\cdot\big((\bm p'_{1}+\bm p_{1})+i(1+\chi_{\ssst{V}})\bm\sigma_{1}\times\bm q_{1}\big)\Big)\Big\}\nonumber \\
 &  & \times\frac{(2\pi)^{3}\delta^{(3)}(\cdots)}{(\bm q_{1}^{2}+m_{\rho}^{2})(\bm q_{2}^{2}+m_{\rho}^{2})}+(1\leftrightarrow2)\,,\label{eq:rho mesonic}\\
\bm j_{mesonic}^{\rho\pi} & = & \frac{e\grhoNN g_{\rho\pi\gamma}h_{\pi}^{1}}{\sqrt{2}m_{\rho}}(\bm\tau_{1}\times\bm\tau_{2})^{z}
[\bm q_{1}\times\bm q_{2}]\frac{(2\pi)^{3}\delta^{(3)}(\cdots)}{(\bm q_{1}^{2}+m_{\rho}^{2})(\bm q_{2}^{2}+m_{\pi}^{2})}+(1\leftrightarrow2)\,,\\
\bm j_{mesonic}^{\omega\pi} & \approx & 0\,,\end{eqnarray}
where $k_{0}=E_{1}+E_{2}-E_{1}'-E_{2}'=E_{i}-E_{f}$. Specially note
that there is a contribution to the charge density at the same order
as the nucleon charge:\begin{eqnarray}
\rho_{mesonic}^{\rho} & = & -2e\grhoNN\left(h_{\rho}^{0}-\frac{h_{\rho}^{2}}{2\sqrt{6}}\right)i(\bm\tau_{1}
\times\bm\tau_{2})^{z}\bm\sigma_{2}\cdot\bm k\frac{(2\pi)^{3}\delta^{(3)}(\cdots)}{(\bm q_{1}^{2}+m_{\rho}^{2})(\bm q_{2}^{2}+m_{\rho}^{2})}+(1\leftrightarrow2)\,.\label{eq:PNC MEC last}\end{eqnarray}

\subsection{PNC $NN$ Interaction and Current Conservation}

By considering the one-boson exchange diagrams where one of the meson-nucleon
couplings is PC and the other PNC, the DDH potential in momentum space
could be expressed as

\begin{eqnarray}
V_{\ssst{PNC}}^{\pi} & = & \frac{\gpiNN h_{\pi}^{1}}{2\sqrt{2}\mN}i\left(\vtau_{1}\times\vtau_{2}\right)^{z}
\left(\vsig_{1}+\vsig_{2}\right)\cdot\bm u_{\pi}\,,\label{eq:vpnc-pi-p}\\
V_{\ssst{PNC}}^{\rho} & = & \frac{-\grhoNN}{\mN}\left[\left(h_{\rho}^{0}\vtau_{1}\cdot\vtau_{2}+
\frac{h_{\rho}^{1}}{2}(\tau_{1}^{z}+\tau_{2}^{z})+\frac{h_{\rho}^{2}}
{2\sqrt{6}}(3\tau_{1}^{z}\tau_{2}^{z}-\vtau_{1}\cdot\vtau_{2})\right)
\right.\nonumber \\
 &  & \left.\times\big(i(1+\chi_{\ssst{V}})\,(\vsig_{1}\times\vsig_{2})\cdot\bm u_{\rho}+(\vsig_{1}-\vsig_{2})\cdot\bm v_{\rho}\big)-\frac{h_{\rho}^{1}}{2}(\tau_{1}^{z}-\tau_{2}^{z})(\vsig_{1}+
\vsig_{2})\cdot\bm v_{\rho}\right]\,,\label{eq:vpnc-rho-p}\\
V_{\ssst{PNC}}^{\omega} & = & \frac{-\gomegaNN}{\mN}\left[\left(h_{\omega}^{0}+\frac{h_{\omega}^{1}}{2}
(\tau_{1}^{z}+\tau_{2}^{z})\right)\big(i(1+\chi_{\ssst{S}})(\vsig_{1}
\times\vsig_{2})\cdot\bm u_{\omega}+(\vsig_{1}-\vsig_{2})\cdot\bm v_{\omega}\big)\right.\nonumber \\
 &  & \left.+\frac{h_{\omega}^{1}}{2}\,(\tau_{1}^{z}-\tau_{2}^{z})(\vsig_{1}
+\vsig_{2})\cdot\bm v_{\omega}\right]\,,\label{eq:vpnc-omega-p}\end{eqnarray}
where\[
\bm u_{\ssst{X}}=\frac{\bm q}{\bm q^{2}+m_{\ssst{X}}^{2}}\,;\quad\bm v_{\ssst{X}}=\frac{(\bm p'_{1}+\bm p_{1})-(\bm p'_{2}+\bm p_{2})}{2\, ( \bm q^{2}+m_{\ssst{X}}^{2} ) }\,,\]
with $\bm q$ denotes the meson 3-momentum.

To prove the conservation of these PNC currents at the operator level,
we showed explicitly the following matrix element identities (with
bra $\langle\bm p'_{1},\bm p'_{2}|$ and ket $|\bm p_{1},\bm p_{2}\rangle$): 

\begin{eqnarray}
\langle[\rho^{(1)}\,,\, V_{\ssst{PNC}}^{\pi}]\rangle & \textrm{=} & \bm k\cdot\langle\bm j_{pair}^{\pi}+\bm j_{mesonic}^{\pi}\rangle\,,\\
\langle[\rho^{(1)}\,,\, V_{\ssst{PNC}}^{\rho}]\rangle & = & \bm k\cdot
\langle\bm j_{pair+\ssst{KR}}^{\rho}+\bm j_{mesonic}^{\rho\,\ssst{(I)}}\rangle\,,\\
\langle[\rho^{(1)}\,,\, V_{\ssst{PNC}}^{\omega}]\rangle & = & \bm k\cdot\langle\bm j_{pair}^{\omega}\rangle\,,\\
\langle[\rho_{mesonic}^{\rho}\,,\, H]\rangle & = & \bm k\cdot\langle\bm j_{mesonic}^{\rho\,\ssst{(II)}}\rangle\,,\label{eq:self conserved}\\
0 & = & \bm k\cdot\langle\bm j_{mesonic}^{\rho\pi}+\bm j_{mesonic}^{\omega\pi}\rangle\,,\label{eq:NB continuity}\end{eqnarray}
 where $H$ is the total Hamiltonian, which is the sum of kinetic
energy ($T$) and both PC and PNC potentials ($V_{\ssst{PC}}$ and
$V_{\ssst{PNC}}$); and the $\rho$ mesonic current, Eq. (\ref{eq:rho mesonic}),
is separated into two parts: (I) is proportional to the vector $(\bm q_{1}-\bm q_{2})$,
and (II) contains the rest. The continuity equality of Eq. (\ref{eq:self conserved})
indicates that $(\rho_{mesonic}^{\rho}\,,\,\bm j_{mesonic}^{\rho\,\ssst{(II)}})$
forms a conserved current not constrained by the DDH potential, which
actually results from the self-conserved $\rho\rho\gamma$ vertex
mentioned above, while the last equality shows the transversality
of NB currents. Obviously, the total PNC EM EC operator, $(\rho_{\ssst{PNC}}^{(2)}\,,\,\bm j_{\ssst{PNC}}^{(2)})$
we construct, Eqs. (\ref{eq:PNC MEC first}-\ref{eq:PNC MEC last}),
satisfies the total current conservation condition\[
[\rho^{(1)}+\rho_{\ssst{PNC}}^{(2)}\,,\, T+V_{\ssst{PC}}+V_{\ssst{PNC}}]=\bm k\cdot(\bm j^{(1)}+\bm j_{\ssst{PC}}^{(2)}+\bm j_{\ssst{PNC}}^{(2)})\,,\]
as long as the two-body PC EC, $(\rho_{\ssst{PC}}^{(2)},\bm j_{\ssst{PC}}^{(2)})$,
is conserved, i.e.,\[
[\rho^{(1)}\,,\, V_{\ssst{PC}}]=\bm k\cdot\bm j_{\ssst{PC}}^{(2)}\,,\]
 ($\rho_{\ssst{PC}}^{(2)}$ is higher order in $1/\mN$). Therefore,
at least for the PNC part, we have every conservation condition met.

For the calculation of AM, we have to use both the DDH potential and
current operators in coordinate space, these expressions could be found
in Appendix \ref{sec:MECs in x}.

\section{Deuteron Anapole Moment \label{sec:d AM}}

\subsection{Determination of the Deuteron Wave Function}

Due to the PNC $NN$ interaction, the deuteron wave function, mainly
a $^{3}S_{1}$ state with some fraction of $^{3}D_{1}$ component,
could have parity admixtures in $^{3}\tilde{P}_{1}$ and $^{1}\tilde{P}_{1}$
channels. The former channel, induced by the isovector part of DDH
potential, is dominated by $\pi$-exchange, while the latter one,
resulting from the isoscalar interaction, is purely through the heavy-meson 
exchange. Therefore, we express the full parity-admixed deuteron
wave function as \begin{eqnarray}
\Psi(\mbfr) & = & \frac{1}{\sqrt{4\pi}r}\left[\left(u(r)+\frac{S_{12}(\hat{\bm r})}{\sqrt{8}}w(r)\right)\zeta_{00}\right.\nonumber \\
 &  & \left.-i\sqrt{\frac{3}{8}}\,(\vsig_{1}+\vsig_{2})\cdot\hat{\mbfr}\,\vto\,
\zeta_{10}+i\frac{\sqrt{3}}{2}\,(\vsig_{1}-\vsig_{2})\cdot\hat{\mbfr}\,\voo\,
\zeta_{00}\right]\,\chi_{1J_{z}}\,,\label{eq:d w.f.}\end{eqnarray}
where $S_{12}(\hat{\bm r})\equiv3\bm\sigma_{1}\cdot\hat{\bm r}\,\bm\sigma_{2}\cdot\hat{\bm r}-\bm\sigma_{1}\cdot\bm\sigma_{2}$
and $\chi$ and $\zeta$ represents spinor and isospinor respectively. 

The Schr\"{o}dinger equation in the center of mass frame is \begin{eqnarray}
H\,\Psi(\mbfr) & = & \left[-\frac{1}{\mN}\left(\frac{1}{r}\frac{\partial^{2}}{\partial\, r^{2}}\, r-\frac{l(l+1)}{r^{2}}\right)+V_{C}(r)+V_{T}(r)\, S_{12}(\hat{\bm r})+V_{\ssst{PNC}}(\mbfr)\right]\,\Psi(\mbfr)\nonumber \\
 & = & E\,\Psi(\mbfr)\,,\end{eqnarray}
 where $V_{C}(r)$ and $V_{T}(r)$ are the central and tensor parts of
the strong potential, respectively, and $V_{\ssst{PNC}}(\mbfr)$ is
the sum of the PNC potentials given above. Up to the linear order
of PNC $NN$ interaction, radial wave functions of each channel satisfy
the differential equations 
\begin{eqnarray}
&&
u''(r)+\mN\,\left(E-V_{C}(r)\right)u(r)=\sqrt{8}\mN V_{T}(r)w(r)\,,\\
&&
w''(r)-\frac{6}{r^{2}}w(r)-\mN\,(E-V_{C}(r)+2V_{T}(r))w(r)=\sqrt{8}\mN V_{T}(r)u(r)\,,\\
&&
\tilde{v}_{3p1}''(r)-\frac{2}{r^{2}}\vto+\mN(E-V_{C}(r)-2V_{T}(r))\vto=
\nonumber \\
&&
\frac{2}{\sqrt{3}}\left[\left(u(r)+\frac{1}{\sqrt{2}}w(r)\right)
\frac{\partial}{\partial r}\left(F_{\pi}^{1}(r)+\sqrt{2}F_{\rho}^{1}(r)-\sqrt{2}F_{\omega}^{1}(r)\right)
\right.
\nonumber \\
&&
\left.+2\sqrt{2}\left(F_{\rho}^{1}(r)-F_{\omega}^{1}(r)\right)\frac{\partial}
{\partial r}\left(u(r)+\frac{1}{\sqrt{2}}w(r)\right)-\frac{2\sqrt{2}}{r}
\left(F_{\rho}^{1}(r)-F_{\omega}^{1}(r)\right)\left(u(r)-\sqrt{2}w(r)
\right)\right]
\,, \label{eq:v3p1}\\
&&
\tilde{v}''_{1p1}-\frac{2}{r^{2}}\voo+\mN(E-V_{C}(r)+4V_{T}(r))\voo=
\nonumber \\
&&
\frac{2}{\sqrt{3}}\left[\left(u(r)-\sqrt{2}w(r)\right)\frac{\partial}
{\partial r}\left(3\chi_{\ssst{V}}F_{\rho}^{0}(r)-\chi_{\ssst{S}}F_{\omega}^{0}(r)\right)
\right.
\nonumber \\
&&
\left.-2\left(3F_{\rho}^{0}(r)-F_{\omega}^{0}(r)\right)\frac{\partial}{\partial r}\left(u(r)-\sqrt{2}w(r)\right)+\frac{2}{r}\left(3F_{\rho}^{0}(r)
-F_{\omega}^{0}(r)\right)\left(u(r)+2\sqrt{2}w(r)\right)\right]
\,,
\label{eq:v1p1}
\end{eqnarray}
where $F_{\pi}^{1}(r)\equiv g_{\pi NN}h_{\pi}^{1}f_{\pi}(r)$, $F_{\rho}^{0}(r)\equiv g_{\rho NN}h_{\rho}^{0}f_{\rho}(r)$,
$F_{\rho}^{1}(r)\equiv g_{\rho NN}h_{\rho}^{1}f_{\rho}(r)$, $F_{\omega}^{0}(r)\equiv g_{\omega NN}h_{\omega}^{0}f_{\omega}(r)$,
and $F_{\omega}^{1}(r)\equiv g_{\omega NN}h_{\omega}^{1}f_{\omega}(r)$.
In our numerical calculations, $\gpiNN=13.45$, $\grhoNN=2.79$, $\gomegaNN=8.37$,
as well as DDH best values (in units of $10^{-7}$): $\hpiNN=4.6$,
$h_{\rho}^{0}=-11.4$, $h_{\rho}^{1}=-0.2$, $h_{\rho}^{2}=-9.5$,
$h_{\omega}^{0}=-1.9$, and $h_{\omega}^{1}=-1.1$, are assumed.

\subsection{Anapole Moment : Expressions of Matrix Elements}

The anapole operator we use takes the form\begin{equation}
\mbfa\equiv\frac{2\pi}{3}\int d\mbfx\;\mbfx\times(\mbfx\times\mbfj(\mbfx))\,,\label{eq:orig-am}\end{equation}
where $\mbfj(\mbfx)$ contains all the currents discussed in Sec.
\ref{sec:MECs}. Note that this form is equivalent to what have been
recommended in Refs. \cite{HaHM89,HLRM02}, a result from implementing
the extended Siegert's theorem \cite{FrFa84}. 

With the deuteron wave function, Eq. (\ref{eq:d w.f.}), we obtain
the anapole moment for the spin term; \begin{eqnarray}
\mbfa_{spin} & = & -\frac{\pi}{\sqrt{6}\,\mN}\left[\mu_{\ssst{V}}\,\int dr\, r\,\left(u(r)-\sqrt{2}w(r)\right)\vto\right.\nonumber \\
 &  & \left.\hspace{2cm}-\sqrt{2}\,\mu_{\ssst{S}}\,\int dr\, r\,\left(u(r)+\frac{1}{\sqrt{2}}w(r)\right)\voo\right]\, e\,\mbfI\,,\end{eqnarray}
 where $\bm I\equiv\frac{1}{2}\chi_{1\ssst{J_{z}}}^{\dagger}(\bm\sigma_{1}
+\bm\sigma_{2})\chi_{1\ssst{J_{z}}}$
is the intrinsic spin taken in the spinor basis, and we note that
this is equivalent to the total angular momentum taken in the total
angular momentum basis, i.e., $\bm I=\langle J=1,J_{z}|\bm J|J=1,J_{z}\rangle$
\footnote{Since the anapole moment is a vector moment, it should be proportional
to the total angular momentum which is the only intrinsic vector of a system. %
}.

The matrix element of the convection current is written as \begin{eqnarray}
\mbfj_{conv}(\mbfx) & = & \mbfj_{conv}^{+}(\mbfx)+\mbfj_{conv}^{-}(\mbfx)\,,\nonumber \\
\mbfj_{conv}^{+}(\mbfx) & \equiv & \frac{e}{4\mN}\int d\mbfr_{1}d\mbfr_{2}\frac{1}{\sqrt{4\pi}r}\chi_{1J_{z}}^{\dagger}\left(u(r)
+\frac{S_{12}(\hat{\mbfr})}{\sqrt{8}}w(r)-i\frac{\sqrt{3}}{2}(\vsig_{1}-\vsig_{2})
\cdot\hat{\mbfr}\tilde{v}_{1p1}(r)\right)\nonumber \\
 &  & \hspace{-0.4cm}\times(\mbfp_{1},\,\,\mbfp_{2})^{+}\left(u(r)
+\frac{S_{12}(\hat{\mbfr})}{\sqrt{8}}w(r)+i\frac{\sqrt{3}}{2}(\vsig_{1}
-\vsig_{2})\cdot\hat{\mbfr}\tilde{v}_{1p1}(r)\right)\frac{1}{\sqrt{4\pi}r}
\chi_{1J_{z}}\,,\\
\mbfj_{conv}^{-}(\mbfx) & \equiv & \frac{e}{4\mN}\int d\mbfr_{1}d\mbfr_{2}\frac{1}{\sqrt{4\pi}r}\chi_{1J_{z}}^{\dagger}\nonumber \\
 &  & \times\left[\left(u(r)+\frac{S_{12}(\hat{\mbfr})}{\sqrt{8}}w(r)\right)(\mbfp_{1},
\,\,\mbfp_{2})^{-}\left(-i\sqrt{\frac{3}{8}}\right)(\vsig_{1}+\vsig_{2})
\cdot\hat{\mbfr}\tilde{v}_{3p1}(r)\right.\nonumber \\
 &  & \left.+i\sqrt{\frac{3}{8}}(\vsig_{1}+\vsig_{2})\cdot\hat{\mbfr}\tilde{v}_{3p1}(r)
(\mbfp_{1},\,\,\mbfp_{2})^{-}\left(u(r)+\frac{S_{12}(\hat{\mbfr})}
{\sqrt{8}}w(r)\right)\right]\frac{1}{\sqrt{4\pi}r}\chi_{1J_{z}}
\,,\label{eq:conv-rel}\end{eqnarray}
 where \begin{eqnarray}
(\mbfp_{1},\,\,\mbfp_{2})^{\pm}\equiv\left\{ \mbfp_{1},\,\,\delta^{3}(\mbfx-\mbfr_{1})\right\} \pm\left\{ \mbfp_{2},\,\,\delta^{3}(\mbfx-\mbfr_{2})\right\} \,.\end{eqnarray}
 As been argued in \cite{HyDe03} \begin{equation}
\frac{2\pi}{3}\int d\mbfx\mbfx\times\left(\mbfx\times\mbfj_{conv}^{-}(\mbfx)\right)=\left\langle -i\frac{\pi e}{12\mN}\left[\mbfl^{2},\,\,\mbfr\right]\right\rangle \,,\end{equation}
 where $\mbfl=\mbfr\times\mbfp$ %
\footnote{We stress that the equality holds for the matrix elements and the
operators themselves do not satisfy the equality.%
}. Indeed, it can be shown that the matrix element of the operator $(\mbfp_{1},\,\,\mbfp_{2})^{+}$
is proportional to \[
-2i\mbfR+\mbfR\left(\mbfR\cdot\mbfP\right)+\mbfr\left(\mbfR\cdot\mbfp
+\frac{1}{4}\mbfr\cdot\mbfP\right)-\left(\mbfR^{2}+\frac{1}{4}\mbfr^{2}\right)
\mbfP-2\mbfR\cdot\mbfr\,\mbfp\,,\]
 where $\mbfR=(\mbfr_{1}+\mbfr_{2})/2$ is the coordinate of the center
of mass and $\mbfP=\mbfp_{1}+\mbfp_{2}$ is its conjugate momentum.
In the center of mass frame where only the relative coordinate or
momentum are relevant, the terms proportional to $\mbfR$ or $\mbfP$
are discarded. Consequently, the contribution from $(\mbfp_{1},\,\,\mbfp_{2})^{+}$
is zero and the magnitude from the convection current is determined
by Eq. (\ref{eq:conv-rel}). The anapole moment contributed by the convection
term then reads \begin{eqnarray}
\mbfa_{conv} & = & \frac{1}{3}\frac{\pi}{\sqrt{6}\,\mN}\int dr\, r\,\left(u(r)-\sqrt{2}w(r)\right)\vto\, e\,\mbfI\,.\end{eqnarray}

Contributions from the PNC exchange currents are evaluated with the
parity-even channels ($^{3}S_{1}$ and $^{3}D_{1}$) in the initial
and final states. These channels are in spin-triplet, $S=1$, and
isospin-singlet, $T=0$, state and for these states we have the spin
and isospin selections rules \begin{eqnarray}
\langle S=1||\,(\vsig_{1}-\vsig_{2})\,||S=1\rangle & = & \langle S=1||\,(\vsig_{1}\times\vsig_{2})\,||S=1\rangle=0\,,\\
\langle T=0||\,(\vtau_{1}\times\vtau_{2})^{z}||T=0\rangle & = & 0\,,\end{eqnarray}
 where the double bar $||$ means the reduced matrix element. It
is easily seen that the matrix elements of $\mbfj_{pair+\ssst{KR}}^{\rho^{\pm}}$,
$\mbfj_{mesonic}^{\rho^{\pm}}$, $\mbfj_{mesonic}^{\rho\pi}$, and
the part proportional to $\vsig_{1}-\vsig_{2}$ in $\mbfj_{pair}^{\rho^{0}}$
and $\mbfj_{pair}^{\omega}$ vanish. Then we are left with $\pi$-pair,
$\pi$-mesonic, and the parts proportional to $\vsig_{1}+\vsig_{2}$
in $\rho^{0}$-pair and $\omega$-pair terms. Pair and mesonic pion contributions
were already calculated in \cite{HyDe03} which are given as
\begin{eqnarray}
\mbfa_{pair}^{\pi} & = & -\frac{\sqrt{2}\pi\gpiNN}{9\mN}\int dr\, r^{2}\, f_{\pi}(r)\,\left(u(r)+\frac{1}{\sqrt{2}}w(r)\right)\left(u(r)-\sqrt{2}w(r)
\right)e\,\mbfI\,\hpiNN\,, \label{eq:pipair}\\
\mbfa_{mesonic}^{\pi} & = & \frac{\sqrt{2}\pi\gpiNN}{3\mN m_{\pi}}\int dr\, r\, f_{\pi}(r)\left(u(r)+\frac{1}{\sqrt{2}}w(r)\right)\nonumber \\
 &  & \times\left[u(r)\left(1-\frac{1}{3}m_{\pi}r\right)-\frac{1}{\sqrt{2}}w(r)\left(1
+\frac{1}{3}m_{\pi}r\right)\right]e\,\mbfI\,\hpiNN\,,
\label{eq:pimesonic}
\end{eqnarray}
 and $\rho^{0}$- and $\omega$-meson contributions read 
\begin{eqnarray}
\mbfa_{pair}^{\rho^{0}} & = & -\frac{2\pi\, g_{\rho\ssst{NN}}}{9\,\mN}\int dr\, r^{2}\, f_{\rho}(r)\,\left(u(r)+\frac{1}{\sqrt{2}}w(r)\right)\left(u(r)
-\sqrt{2}w(r)\right)\, e\,\mbfI\, h_{\rho}^{1}\,, \label{eq:rhopair}\\
\mbfa_{pair}^{\omega} & = & \frac{2\pi\, g_{\omega\ssst{NN}}}{9\,\mN}\int dr\, r^{2}\, f_{\omega}(r)\,\left(u(r)+\frac{1}{\sqrt{2}}w(r)\right)\left(u(r)
-\sqrt{2}w(r)\right)\, e\,\mbfI\, h_{\omega}^{1}\,.
\label{eq:omegapair}
\end{eqnarray}

\subsection{Numerical Results for the Argonne $v_{18}$  $NN$ Interaction Model}

With the A$v_{18}$ model, we obtain the numerical results (in units of fm$^2$)
\begin{eqnarray}
\mbfa_{spin} & = &  -0.547\, h_{\pi}^{1}\, e\, \bm{I}
+ (-3.7\, h_{\rho}^{1}+ 10.8\, h_{\omega}^{1}+7.3\, h_{\rho}^{0}
+ 3.7\, h_{\omega}^{0}) \times 10^{-3} \, e\,\mbfI\,\,,\\
\mbfa_{conv} & = & 0.039\, h_{\pi}^{1}\, e\, \mbfI
+ (2.7\, h_{\rho}^{1}-7.6\, h_{\omega}^{1})\times 10^{-4}\, e\,\mbfI\,,\\
\mbfa_{ex}^{\pi} & = & \mbfa_{pair}^{\pi}+\mbfa_{mesonic}^{\pi}=(-0.027+0.028)\, h_{\pi}^{1}\, e\,\mbfI\,,\\
\mbfa_{ex}^{\rho} & = & \mbfa_{pair}^{\rho^{0}}=-0.7\times10^{-4}\, h_{\rho}^{1}\, e\,\mbfI\,,\\
\mbfa_{ex}^{\omega} & = & \mbfa_{pair}^{\omega}=1.8\times10^{-4}\, h_{\omega}^{1}\, e\,\mbfI\,.\end{eqnarray}

The contribution from the nucleonic AM, $\bm a_{N}$, should also be
included in the full result. Since the deuteron is an isosinglet state,
only the isoscalar component of the nucleonic AM contributes, i.e.,\begin{equation}
\bm a_{\ssst{N}}=\bra{d}\,\sum_{i=1}^{2}\,(a_{\ssst{S}}^{(1)}+a_{\ssst{V}}^{(1)}
\,\tau_{i}^{z})\,\bm\sigma_{i}\,\ket{d}=2\, a_{\ssst{S}}^{(1)}\,(1-\frac{3}{2}P_{\ssst{D}})\,\bm I\,,\end{equation}
where $a_{\ssst{S.V}}^{(1)}$ denote the isoscalar and isovector nucleonic
AM, and $P_{\ssst{D}}$ is the probability of deuteron $D$-state.
Several theoretical estimates for the nucleonic AM exist \cite{HaHM89,MuHo90,MuHo91,KaSa93,MavK00,Zh..00},
and here we use the result of Ref. \cite{Zh..00} because it is the
most recent one which includes the full DDH interaction at the nucleon
level.

For the pion sector, Ref. \cite{Zh..00} gave

\[
a_{\ssst{S}}^{\pi\,(1)}=
-\frac{g_{\ssst{A}}\, h^1_\pi}{12\sqrt{2}f_{\pi}m_{\pi}}
\frac{\Lambda_{\chi}^{2}}{\mN^{2}}\,(1-\frac{6}{\pi}\frac{m_{\pi}}{\mN}
\ln\frac{\Lambda_{\chi}}{m_{\pi}})\,,\]
where $\Lambda_{\chi}$ is the chiral symmetry breaking scale (the
authors chose it to be $4\pi f_{\pi}$) %
\footnote{Note that a factor of $4\pi/\mN^{2}$ is needed for converting the
dimensionless AM defined in Lorentz-Heaviside units into our
definition.%
}. When setting $\Lambda_{\chi}=\mN$, the leading term is equivalent
to what has been used in Refs. \cite{SaSp98,SaSp01,HyDe03} for the
deuteron AM calculations, while the full result is the same as been
used in Ref. \cite{KhKo00}. By including the heavy mesons, the numerical
result is \cite{Zh..00}
\begin{equation}
a_{\ssst{S}}^{(1)}=-0.274\, h_{\pi}^{1}-0.419\, h_{\rho}^{1}-0.129\, h_{\omega}^{0}\,,
\label{eq:aN}
\end{equation}
and this leads to the nucleonic contribution\begin{equation}
\bm a_{\ssst{N}}=( -0.250\, h_{\pi}^{1}-0.383\, h_{\rho}^{1}-0.118\, h_{\omega}^{0})\, e\, \mbfI \,.\end{equation}

Although we do not have the exact PC EC corresponding to A$v_{18}$
potential, we try to approximate this by the EC coming out from one-pion
exchange diagrams. Corresponding currents 
in the configuration space are given in the Appendix A.
Pair- and mesonic-term of the PC EC contributions read
\begin{eqnarray}
\mbfa^{\ssst{PC}}_{pair} &=&
\frac{1}{3 \sqrt{6}}\left(\frac{g_{\pi \ssst{NN}}}{2 \mN} \right)^2 
\int dr\, {\rm e}^{-m_\pi r}\, (1 + m_\pi r)\, \tilde{v}_{3p1}(r)
\left( u(r) -  \sqrt{2} w(r)\right) \, e\, \mbfI, \\
\mbfa^{\ssst{PC}}_{mesonic} &=&
- \frac{1}{3 \sqrt{6}}\left(\frac{g_{\pi\ssst{NN}}}{2 \mN} \right)^2 
\int dr\, {\rm e}^{-m_\pi r}\, \tilde{v}_{3p1}(r)
\left[ 3 u(r) - m_\pi\, r
\left(u(r) + \frac{1}{\sqrt{2}} w(r)\right) \right] e\, \mbfI.
\end{eqnarray}
Numerically we obtain
\begin{eqnarray}
\bm a_{ex}^{\ssst{PC}} &=& 
\bm{a}^{\ssst{PC}}_{pair} 
+ \bm{a}^{\ssst{PC}}_{mesonic} \nonumber \\ 
&=& - (7.8\ h^1_\pi + 0.4\ h^1_\rho - 1.1\ h^1_\omega)
\times 10^{-4}\, e\, \bm{I}.
\end{eqnarray}
In the next section, we will discuss if this is a reasonable approximation
and to what extent is the breaking of current conservation due to
the non-exact PC EC.

Finally, the full deuteron AM could be expressed as:
\begin{eqnarray}
\bm a_{d} & = & \bm a_{spin}+\bm a_{conv}+\bm a_{ex}^{\ssst{PNC}}+\bm a_{\ssst{N}}+\bm a_{ex}^{\ssst{PC}} \nonumber \\
&=& ( -0.756\, h^1_\pi - 0.387\, h^1_\rho + 0.010\, h^1_\omega
+ 0.007\, h^0_\rho - 0.114\, h^0_\omega )\, e\, \mbfI.
\end{eqnarray}

\section{Discussions \label{sec:dis}}

\begin{figure}
\begin{center}
\includegraphics[scale=0.40]{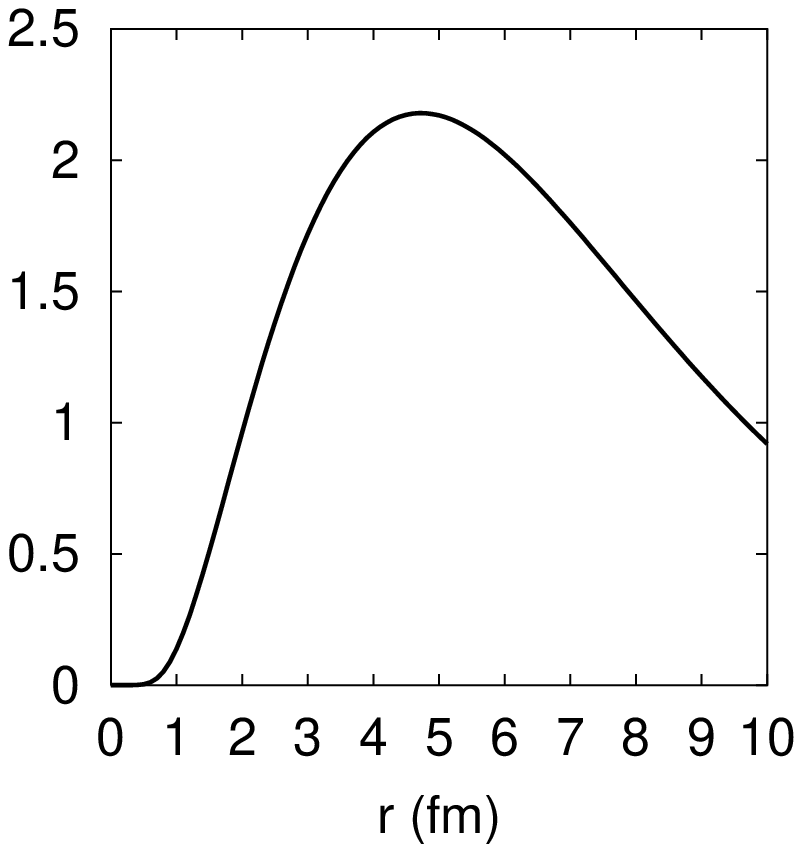}
\includegraphics[scale=0.40]{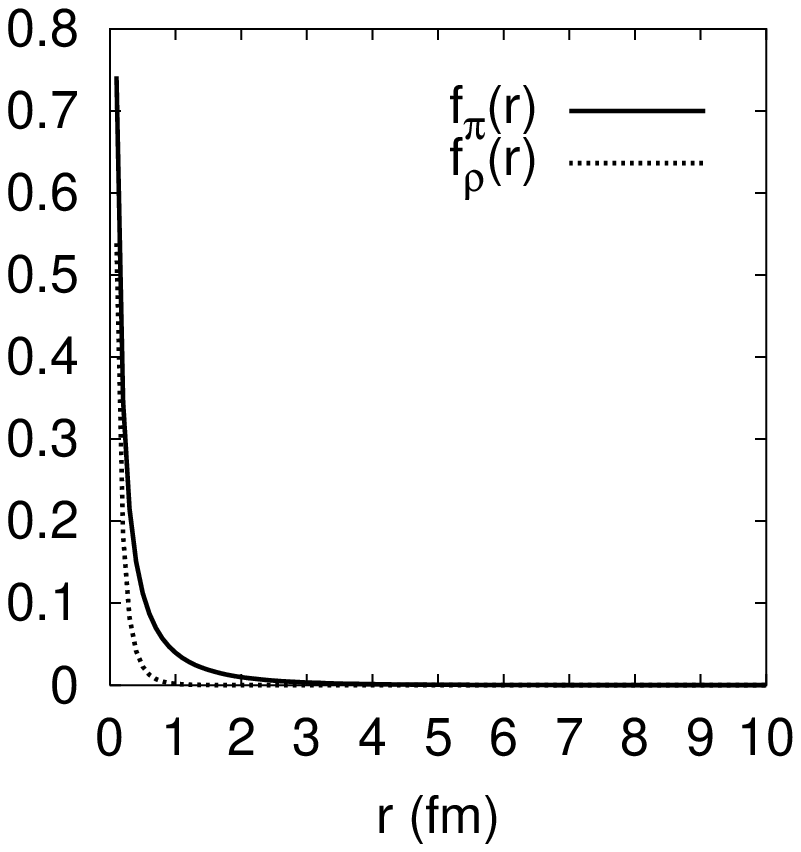}
\includegraphics[scale=0.40]{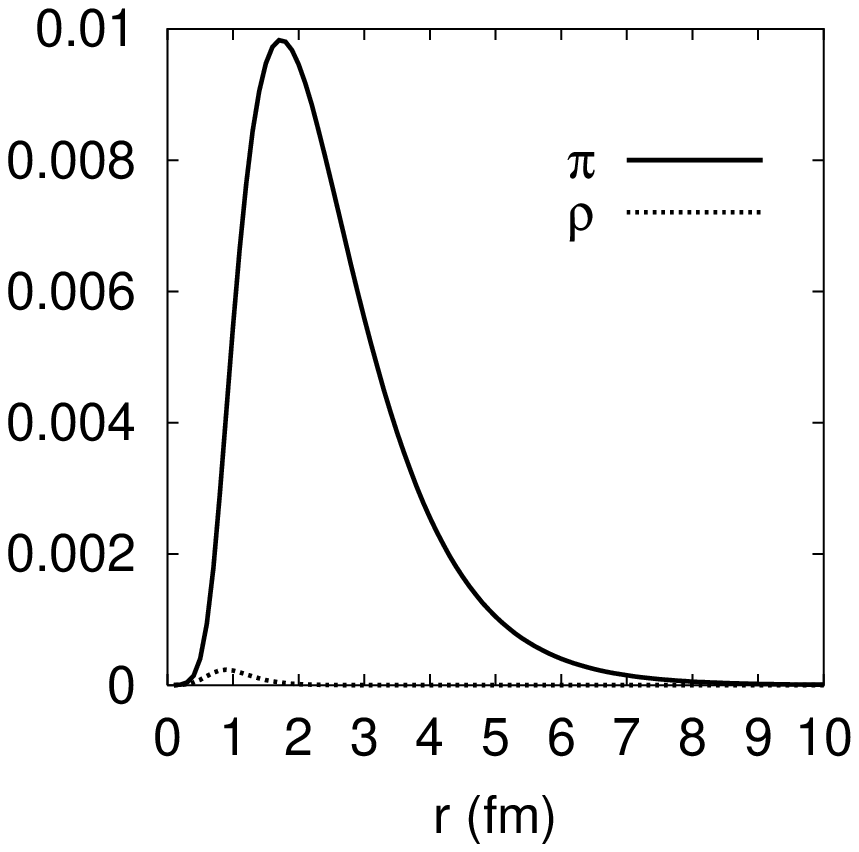}
\end{center}
\caption{The left panel shows the behavior of
$r^2\, (u(r)+w(r)/\sqrt{2})(u(r) - \sqrt{2} w(r))$. The central panel 
compares the Yukawa function of the pion (solid line) and rho-meson 
(dotted line). The right panel is the behavior of the integrand for the $\pi-$
and $\rho-$ pair terms in Eq. (\ref{eq:pipair}) and Eq. (\ref{eq:rhopair}), 
respectively
\label{fig:suppression}
}
\end{figure}

Presenting the results, we separated the contributions of the pion 
and the heavy mesons so that their relative order of magnitude can be compared
easily. In the result of the nuclear part, i.e. the contributions from
deuteron wave function or exchange currents, which include spin, convection,
PNC EC and PC EC, the magnitudes of the heavy meson contributions are 
commonly suppressed by an order of 2 or 3.
This suppression can be understood most easily from 
Eqs. (\ref{eq:pipair}, \ref{eq:rhopair}, \ref{eq:omegapair}), 
PNC pair terms of $\pi$, $\rho$ and $\omega$ mesons, respectively.
Aside from the weak coupling constants, they differ only by the Yukawa 
functions $f_\pi(r)$ or $f_{\rho,\, \omega}(r)$ in the integrand.
At small $r$, the deuteron wave function is proportional to $r$ (for the
simplicity of argument, we neglect $D-$state wave function) and 
increases very fast up to $r \sim 2$ fm at which the maximum is reached.
Afterwards, the wave function decreases very slowly, converging to zero.
The factor $r^2$ multiplied by $u^2(r)$ gives rise to the suppression at 
short range. The left panel of Fig. \ref{fig:suppression} shows 
this behavior. One can expect that this $r^4$ behavior at short distances makes the contribution from $r \leq 1$ fm quite small.
On the contrary, the Yukawa function, which behaves like $1/r$ at small
$r$ and also depends strongly on the mass of the meson, is short-range peaked.
A comparison of $f_\pi(r)$ and $f_\rho(r)$ is given in the central panel 
of Fig. \ref{fig:suppression}. The quantity
$f_\rho(r)$ is non-negligible for about $r \leq 0.5$ fm but the remaining
part of the integrand, approximately $r^2\, u^2(r)$ is small in this region.
Consequently the heavy meson will be suppressed 
substantially compared to $\pi$.
The right panel of Fig. \ref{fig:suppression} shows the behavior of the total
integrands in Eqs. (\ref{eq:pipair}, \ref{eq:rhopair}). 
If we approximate the integrand  as $r^3\, \exp(- m_{\ssst{X}}r)$, the 
maximum of the integrand is reached at $r \sim 3/m_{\ssst{X}}$.
This approximation is good for $\rho$ but bad for $\pi$, however, this can 
provide a semi-quantitative explanation for the suppression of the heavy 
mesons. At the maximum, the integrand has the value proportional to 
$1/m^3_{\ssst{X}}$. 
If comparison is made for $\pi$ and $\rho$, the maximum value of 
$\rho$ is smaller than $\pi$ by a factor of 1/200, which can account for
the suppression of order 2 of the heavy mesons.

To a lesser extent, a similar argument can be applied to the qualitative 
understanding of the heavy-meson suppression of spin and convection terms. 
The right hand side of Eqs. (\ref{eq:v3p1}, \ref{eq:v1p1}) 
which are the sources of parity admixed $P-$states also contains 
the Yukawa functions. When the equations are solved, i.e. integrated
with respect to $r$, one can expect some amount of suppression
for the heavy mesons as was discussed in the analysis of pair EC's, 
but without the factor $r^2$.
In Tab. \ref{tab:sum1}, we summarize the ratio of the heavy meson 
contribution to the pion with the weak coupling constants given by DDH 
best values. The magnitudes of the heavy meson terms are suppressed by
an order of two compared to the pion terms in common. As far as the nuclear part 
is concerned, $\pi$ contribution is by far dominant. If one can 
disentangle the nuclear part contribution from the total deuteron anapole 
moment, this can provide information to extract the magnitude 
of $h^1_\pi$ with high accuracy.
As a side remark, it is noticed that the combination of the isovector 
couplings, $h^1_{\pi}$, $h^1_{\rho}$ and $h^1_{\omega}$, entering the spin and 
convection contributions is close to the one determining low energy PNC  
nucleon-nucleon scattering, roughly 
$3\;h^1_{\pi} +0.02\;h^1_{\rho}-0.06\;h^1_{\omega}$ \cite{Desp98}. 
This lets us think that the Danilov's approach \cite{Dani65}, whose application for 
the deuteron anapole moment was proposed later on by Savage \cite{Sa01}, should 
provide a reasonable estimate for this contribution.

\begin{table}
\begin{center}
\begin{tabular}{|c||c|c|c|c|c||c|}\hline
term & spin & conv. & PNC EC & PC EC & nucleonic & total \\ \hline
$(\rho+\omega)/\pi$ & 4.0 & 0.6 & $- 2.0$ & $- 3.1$ & $- 26.2$ & $-5.9$ \\
\hline
\end{tabular} 
\end{center}
\caption{Comparison of the heavy meson with the pion contribution to 
the deuteron anapole moment, term by term. The ratios are given in \%.
\label{tab:sum1}
}
\end{table}

Contrary to the nuclear part, the nucleonic one has sizable contribution
from the heavy mesons.
The relative ratio of the heavy meson is about 26\% of the pion based on the nucleonic
anapole estimate of Ref. \cite{Zh..00}. However, we should also note that these authors
considered more non-DDH type couplings such as non-Yukawa type $\pi NN$ couplings and the 
inclusion of hyperons. So far, no detailed knowledge about these exotic couplings exist, and
the theoretical uncertainty could be huge. The study of these terms and their implication
for two-body nuclear contributions will be an interesting topic for further exploration.

Another issue that should be addressed is the gauge invariance of the results.
We showed in Sect. \ref{sec:MECs} that the PNC ECs we constructed satisfy
the gauge invariance with the DDH potential.
However, with the phenomenological strong interaction models like the one 
adopted in this work, gauge invariance may not be 
satisfied if the ECs are not consistent with the phenomenological
potentials. For A$v_{18}$, the potential contains 18 types of operators but
we took into account the dominant ECs only, which naturally leads to 
inconsistency, or breakdown of gauge invariance.
Even though the inconsistency may only change the result
slightly, the investigation of the extent to which gauge invariance is broken could help to get an estimate of the error.

Most phenomenological potentials are very complicated and the analytic 
analysis
of gauge invariance is a formidable or practically impossible work.
However as suggested in \cite{HyDe03}, one can estimate the amount of 
gauge invariance breaking by comparing the anapole moment obtained from 
Eq. (\ref{eq:orig-am}) and 
\begin{equation}
\bm{a} = - \pi \int d\bm{x}\, x^2 \bm{j}(\bm{x})\, ,
\label{eq:alt-am}
\end{equation}
which can be obtained from Eq. (\ref{eq:orig-am}) with the assumption
\begin{equation}
\nabla \cdot \bm{j}(\bm{x}) = 0\, . 
\label{eq:crnt-conserv}
\end{equation} 
In \cite{HyDe03}, we observed that the inclusion of the PC ECs of $\pi$
removes almost all the inconsistency. In this work, we follow the 
same procedure to investigate the breakdown of gauge invariance.

The spin current can be shown easily to satisfy Eq. (\ref{eq:crnt-conserv}).
The convection, pair and mesonic anapole moment of $\pi$, $\rho-$ and 
$\omega-$mesons with the definition given by Eq. (\ref{eq:alt-am})
can be calculated straightforwardly to give
\begin{eqnarray}
\bm{a}^{\ssst{CC}}_{conv} &=& \frac{\pi}{2 \sqrt{6}\, \mN}
\int dr\, r^2\, \left[
u(r) \left(\tilde{v}'_{3p1}(r) + 2 \frac{\tilde{v}_{3p1}(r)}{r} \right)
+ \frac{w(r)}{\sqrt{2}}
\left(\tilde{v}'_{3p1}(r) - \frac{\tilde{v}_{3p1}(r)}{r} \right) \right]
\, e\, \mbfI,\\
\bm{a}^{\pi,\ssst{CC}}_{pair} &=& 
- \frac{\pi\, g_{\pi\, \ssst{NN}}}{2 \sqrt{2}\, \mN}
\int dr\, r^2\, f_\pi(r) \left( u^2(r) - \frac{w^2(r)}{2} \right)\, 
h^1_\pi\, e\, \mbfI, 
\\
\mbfa^{\pi,\ssst{CC}}_{mesonic} &=& 
\frac{g_{\pi\ssst{NN}}}{24\sqrt{2}\, \mN\, m_\pi} 
\int dr\, {\rm e}^{-m_\pi r}  \nonumber \\
& & \times \left[
\left( u^2(r) - \frac{w^2(r)}{2}\right) (m_\pi r + 4) - 
\frac{1}{3}\left(u(r) + \frac{w(r)}{\sqrt{2}} \right)^2
m_\pi r\, (m_\pi r + 3) \right] h^1_\pi\, e\, \mbfI, \\
\bm{a}^{\rho^0,\ssst{CC}}_{pair} &=&
- \frac{\pi\, g_{\rho\, \ssst{NN}}}{2 \, \mN} \int dr\, r^2\, 
f_\rho(r) \left( u^2(r) - \frac{w^2(r)}{2} \right)\, h^1_\rho\, e\, \mbfI,
\\
\bm{a}^{\omega,\ssst{CC}}_{pair} &=&
\frac{\pi\, g_{\omega\, \ssst{NN}}}{2 \, \mN} \int dr\, r^2\,
f_\omega(r) \left( u^2(r) - \frac{w^2(r)}{2} \right)\, h^1_\omega\, e\, \mbfI,
\\
\mbfa^{\ssst{PC},\ssst{CC}}_{pair} &=& \frac{1}{\sqrt{6}}
\left(\frac{g_{\pi\ssst{NN}}}{2\, \mN} \right)^2
\int dr\, {\rm e}^{-m_\pi r} (1+ m_\pi r)\,
 \tilde{v}_{3p1}(r)\left(u(r) - \frac{1}{\sqrt{8}}\, w(r)\right)
\, e\, \mbfI\, , \\
\mbfa^{\ssst{PC},\ssst{CC}}_{mesonic} &=& - \frac{1}{12 \sqrt{6}}
\left(\frac{g_{\pi\ssst{NN}}}{2\, \mN} \right)^2
\int dr\, {\rm e}^{-m_\pi r}\, \tilde{v}_{3p1}(r) \nonumber \\
& & \times \left[ u(r) (18 + 2 m_\pi r - m^2_\pi r^2) - 
\frac{w(r)}{\sqrt{2}}\, m_\pi r\, (4 + m_\pi r) \right]\, e\, \mbfI,
\end{eqnarray}
where the superscript $CC$ denotes the quantity calculated with 
Eq. (\ref{eq:alt-am}).
Numerical results are summarized in Tab. \ref{tab:sum2}.

\begin{table}
\begin{center}
\begin{tabular}{|c|c|c|c||c|c|c|}\hline
 & \multicolumn{3}{c||} {Eq. (\ref{eq:orig-am})} &
   \multicolumn{3}{c|} {Eq. (\ref{eq:alt-am})} \\ \cline{2-7}
 & $h^1_\pi$ & $h^1_\rho$ & $h^1_\omega$ &
$h^1_\pi$ & $h^1_\rho$ & $h^1_\omega$ \\ \hline
conv.             & 0.03925 & 0.00027 & $-0.00076$& 
                    0.05348 & 0.00024 & $-0.00066$ \\ \hline
PNC $\pi$ pair    &$-0.02668$ &         & 
                  &$-0.07895$ &         &        \\ \hline
PNC $\pi$ mesonic & 0.02830 &         & 
                  & 0.04706 &         &        \\ \hline
PNC $\rho$ pair   &         &$-0.00007$ & 
                  &         &$-0.00023$ &         \\ \hline
PNC $\omega$ pair &         &         & 0.00018
                  &         &         & 0.00061 \\ \hline
PC  $\pi$ pair    & 0.01400 & 0.00013 &$-0.00038$
                  & 0.06325 & 0.00062 &$-0.00177 $ \\ \hline
PC $\pi$ mesonic  &$-0.01478$&$-0.00017$& 0.00049
                  &$-0.04366$&$-0.00047$ & 0.00133  \\ \hline\hline
Total$-$PC          & 0.04087 & 0.00020 &$-0.00058$
                  & 0.02159 & 0.00001 &$-0.00005$ \\ \hline\hline
Total             & 0.04010 & 0.00016 &$-0.00047$
                  & 0.04117 & 0.00017 &$-0.00049 $\\ \hline\hline
\end{tabular}
\end{center}
\caption{Coefficients of the weak coupling constants for given terms
with different definitions of the anapole operator, Eq. (\ref{eq:orig-am})
and Eq. (\ref{eq:alt-am}).
\label{tab:sum2}
}
\end{table}

Judging from the "Total" column in Table II, two definitions of the anapole
operator only differ in results by 3\%, 6\%, and 4\% for the $\pi$-, $\rho$-, and
$\omega$-exchange respectively; this means our result satisfy current
conservation very well. This is quite surprising because only the PC one-pion 
exchange current---not fully consistent with the adopted strong potential, 
A$v_{18}$---is included in our calculation. The reason is partly because the AM is a 
$r^2$-weighted moment, plus the deuteron wave function peaks around 2 fm with a long 
tail; therefore the long-range physics, which is dominated by the one-pion 
exchange including the case of A$v_{18}$, becomes much more important. This explains 
why our conclusion differs from the one found in Ref. \cite{Scetal03} 
where large effects due to undetermined ECs were found.  
Their observation was concerning a different operator (E1 transition), 
which has a  range relatively shorter than the quantity considered 
in this work.

Furthermore, one can observe that for the contributions from exchange currents,
Eq. (\ref{eq:orig-am}) always give smaller values than Eq. (\ref{eq:alt-am}). This 
implies the calculation using Eq. (\ref{eq:orig-am}) suffers less from the 
incomplete knowledge or uncertainty of exchange currents which causes the 
breaking of current conservation. For example, if the PC pion ECs are left out 
in our calculation, by comparing the "Total" and "Total-PC" columns in Table 
II, the error is 2\%, 25\%, and 23\% for $\pi$, $\rho$, and $\omega$ component 
respectively when Eq. (\ref{eq:orig-am}) is used; while the error is -46\%, -94\%, 
and -90\% for Eq. (\ref{eq:alt-am}). 
Perhaps, the reason for the small contribution of ECs in the former case 
is to be looked for in the proportionality of the dominant ECs to the
position vector $\vec{x}$, which readily gives zero when inserted into the  
corresponding anapole operator, Eq. (\ref{eq:orig-am}). Therefore the merit of 
Eq. (\ref{eq:orig-am}) is justified, at least for the case of deuteron.

In conclusion, we constructed the PNC ECs due to one $\pi-$, $\rho-$ and 
$\omega-$exchange, and showed that they satisfy the current conservation; and hence 
are fully consistent with the adopted DDH PNC potential.
An application was made to the calculation of the deuteron anapole moment.
We observed that, for the nuclear part, the contribution of heavy mesons is 
suppressed by an order of two compared to the pion, a result consistent with the 
similar work by Blunden \cite{Blun02}. 
Consistency with A$v_{18}$ was also checked. We found that the approximation of 
using only the PC one-pion exchange current is pretty good---the breaking of current 
conservation only amounts to a few percent, and this should be fixed by using the 
PC exchange current fully consistent with A$v_{18}$.
Therefore, the contribution from the nuclear part to the deuteron anapole moment can 
be determined with an error less than 5\%, while the major uncertainty should come 
from the nucleonic anapole moment instead.

\section*{Acknowledgements}
Work of C.H.H. is partially supported by KOSEF (Grant No.
2000-2-11100-004-4).

\appendix

\section{PNC $NN$ Potential and Exchange Currents in Coordinate Space \label{sec:MECs in x}}

The DDH potential in coordinate space could easily be obtained by
applying the following transformation rules to Eqs. (\ref{eq:vpnc-pi-p}-\ref{eq:vpnc-omega-p})\begin{eqnarray*}
\bm u_{\ssst{X}} & \rightarrow & \bm u_{\ssst{X}}(\bm r)=[\bm p\,,\, f_{\ssst{X}}(r)]\,,\\
\bm v_{\ssst{X}} & \rightarrow & \bm v_{\ssst{X}}(\bm r)=\{\bm p\,,\, f_{\ssst{X}}(r)\}\,,\end{eqnarray*}
where $\bm r\equiv\bm r_{1}-\bm r_{2}$; $r=|\bm r|$; $\bm p\equiv(\bm p{}_{1}-\bm p{}_{2})/2=-i\bm\nabla_{r}$;
and the Yukawa functions $f_{\ssst{X}}(r)$ are defined as\[
f_{\ssst{X}}(r)=\frac{{\rm e}^{-m_{\ssst{X}}r}}{4\pi r}\,.\]

For the PNC MECs, we list all the leading-order, $O(1/\mN)$, 3-currents,
which are relevant for the AM calculation:

\begin{eqnarray}
\mbfj_{pair}^{\pi}(\mbfx;\,\mbfr_{1},\,\mbfr_{2}) & = & -\frac{e\;\gpiNN\; h_{\pi}^{1}}{2\,\sqrt{2}\;\mN}\;(\vtau_{1}\cdot\vtau_{2}-\tau_{1}^{z}\;
\tau_{2}^{z})\; f_{\pi}(r)\sum_{i=1}^{2}\,\delta^{(3)}(\mbfx-\mbfr_{i})\,\vsig_{i}\,,\\
\mbfj_{mesonic}^{\pi}(\mbfx;\,\mbfr_{1},\,\mbfr_{2}) & = & -\frac{e\;\gpiNN\; h_{\pi}^{1}}{2\,\sqrt{2}\;\mN}\;(\vtau_{1}\cdot\vtau_{2}-\tau_{1}^{z}\;
\tau_{2}^{z})\nonumber \\
 &  & \times(\bm\nabla{}_{1}-\bm\nabla_{2})\;\left[(\vsig_{1}\cdot\bm\nabla{}_{1}
-\vsig_{2}\cdot\bm\nabla{}_{2})\,,\; f_{\pi}(r_{x1})\, f_{\pi}(r_{x2})\right]\,,\\
\mbfj_{pair+\ssst{KR}}^{\rho^{\pm}}(\mbfx;\,\mbfr_{1},\,\mbfr_{2}) & = & -\frac{e\;\grhoNN}{2\,\mN}\, f_{\rho}(r)\left(h_{\rho}^{0}-\frac{1}{2\sqrt{6}}\; h_{\rho}^{2}\right)\nonumber \\
 &  & \hspace{-2cm}\times\Bigg((\vtau_{1}\cdot\vtau_{2}-\tau_{1}^{z}\tau_{2}^{z})
\;(\vsig_{1}-\vsig_{2})\;\Big(\delta^{(3)}(\mbfx-\mbfr_{1})-\delta^{(3)}(\mbfx
-\mbfr_{2})\Big)\nonumber \\
 &  & \hspace{1cm}+(1+\chi_{\ssst{V}})\,(\vtau_{1}\times\vtau_{2})^{z}\,
(\vsig_{1}\times\vsig_{2})\,\sum_{i}\delta^{(3)}(\mbfx-\mbfr_{i})\Bigg)\,,\\
\mbfj_{pair}^{\rho^{0}}(\mbfx;\,\mbfr_{1},\,\mbfr_{2}) & = & -\frac{e\;\grhoNN}{2\,\mN}\, f_{\rho}(r)\;\;\tau_{1}^{z}\;\tau_{2}^{z}\nonumber \\
 &  & \hspace{-2cm}\times\Bigg(\Big(h_{\rho}^{0}+\frac{1}{2}\; h_{\rho}^{1}\;(\tau_{1}^{z}+\tau_{2}^{z})+\frac{1}{\sqrt{6}}\; h_{\rho}^{2}\Big)\;(\vsig_{1}-\vsig_{2})+\frac{1}{2}\; h_{\rho}^{1}\;(\tau_{1}^{z}-\tau_{2}^{z})\;(\vsig_{1}+\vsig_{2})\Bigg)\nonumber \\
 &  & \times\;\Big((1+\tau_{1}^{z})\;\delta^{(3)}(\mbfx-\mbfr_{1})-(1+\tau_{2}^{z})
\;\delta^{(3)}(\mbfx-\mbfr_{2})\Big)\,,\\
\mbfj_{mesonic}^{\rho^{\pm}}(\mbfx;\,\mbfr_{1},\,\mbfr_{2}) & = & -\frac{e\;\grhoNN}{2\,\mN}\left(h_{\rho}^{0}-\frac{1}{2\sqrt{6}}\; h_{\rho}^{2}\right)\,(\vtau_{1}\times\vtau_{2})^{z}\nonumber \\
 &  & \hspace{-2cm}\times\Bigg\{(\bm\nabla_{1}-\bm\nabla{}_{2})\;\bigg(-i\,(\vsig_{1}
-\vsig_{2})\cdot\{(\bm\nabla_{1}-\bm\nabla{}_{2})\,,\, f_{\rho}(r_{x1})\, f_{\rho}(r_{x2})\}\nonumber \\
 &  & \hspace{0.8cm}+(1+\chi_{\ssst{V}})\,(\vsig_{1}\times\vsig_{2})\cdot
\left[(\bm\nabla_{1}-\bm\nabla_{2})\,,\, f_{\rho}(r_{x1})\, f_{\rho}(r_{x2})\right]\bigg)\nonumber \\
 &  & \hspace{-1.5cm}+2\,\nabla_{x}^{a}\;\bigg(i\,\{\nabla_{1}^{a}\,\bm\sigma_{2}
-\nabla_{2}^{a}\,\bm\sigma_{1}+\sigma_{1}^{a}\,\bm\nabla_{2}-\sigma_{2}^{a}
\,\bm\nabla_{1}\,,\, f_{\rho}(r_{x1})\, f_{\rho}(r_{x2})\}\nonumber \\
 &  & \hspace{0cm}-(1+\chi_{\ssst{V}})\;[(\sigma_{1}\times\nabla_{1})^{a}\,
\bm\sigma_{2}-(\sigma_{2}\times\nabla_{2})^{a}\,\bm\sigma_{1}\nonumber \\
 &  & \hspace{2.0cm}+\sigma_{1}^{a}\,\bm\sigma_{2}\times\bm\nabla_{2}-\sigma_{2}^{a}
\,\bm\sigma_{1}\times\bm\nabla_{1}\,,\, f_{\rho}(r_{x1})\, f_{\rho}(r_{x2})]\bigg)\nonumber \\
 &  & \hspace{-1.5cm}+4\, i\,\mN\,(\bm\sigma_{1}+\bm\sigma_{2})\;[H\,,\, f_{\rho}(r_{x1})\, f_{\rho}(r_{x2})]\Bigg\}\\
\mbfj_{pair}^{\omega}(\mbfx;\,\mbfr_{1},\,\mbfr_{2}) & = & -\frac{e\;\gomegaNN}{2\,\mN}\, f_{\omega}(r)\nonumber \\
 &  & \hspace{-2cm}\times\Bigg(\Big(h_{\omega}^{0}+\frac{1}{2}\; h_{\omega}^{1}\;(\tau_{1}^{z}+\tau_{2}^{z})\Big)(\vsig_{1}-\vsig_{2})
+\frac{1}{2}\; h_{\omega}^{1}\;(\tau_{1}^{z}-\tau_{2}^{z})\;(\vsig_{1}+\vsig_{2})
\Bigg)\nonumber \\
 &  & \times\Big((1+\tau_{1}^{z})\;\delta^{(3)}(\mbfx-\mbfr_{1})-(1+\tau_{2}^{z})
\;\delta^{(3)}(\mbfx-\mbfr_{2})\Big)\,,\\
\mbfj_{mesonic}^{\rho\pi}(\mbfx;\,\mbfr_{1},\,\mbfr_{2}) & = & -\frac{e\;\grhoNN g_{\rho\pi\gamma\;}h_{\pi}^{1}}{\sqrt{2}\, m_{\rho}}\,(\bm\tau_{1}\times\bm\tau_{2})^{z}\nonumber \\
 &  & \hspace{-2cm}\times(\bm\nabla_{1}\times\bm\nabla_{2})\,\Big(f_{\rho}(r_{x1})\, f_{\pi}(r_{x2})+f_{\pi}(r_{x1})\, f_{\rho}(r_{x2})\Big), 
\end{eqnarray}
where $r_{xi}\equiv\left|\bm x-\bm r_{i}\right|$; $\bm\nabla_{i}$
and $\bm\nabla_{x}$ act on the source point $\bm r_{i}$ and field point
$\bm x$ respectively; and the superscript $a$ is the index to be
summed from 1 to 3. Note that we separate charged and neutral $\rho$
mesons according to their isospin structure; and the last term of Eq. (A5) should be
combined with a charge density corresponding to Eq. (24) in order to insure the current
conservation.

PC one-pion ECs of the pair and mesonic terms that contribute to the AM read
\begin{eqnarray}
\mbfj^{\ssst{PC}}_{pair}(\mbfx ;\, \mbfr_1,\, \mbfr_2) &=&
\left(\frac{g_{\pi \ssst{NN}}}{2 \mN} \right)^2 (\bm{\tau}_1 \times \bm{\tau}_2)^z
\nonumber \\ & & \times
\Big[ \bm{\sigma}_2\ \delta^{(3)}(\mbfx - \mbfr_2)\ \bm{\sigma}_1 \cdot \nabla
+ \bm{\sigma}_1 \ \delta^{(3)}(\mbfx - \mbfr_1)\ \bm{\sigma}_2  \cdot \nabla
\Big] f_\pi(r), \\
\mbfj^{\ssst{PC}}_{mesonic}(\mbfx ;\, \mbfr_1,\, \mbfr_2) &=&
- \left(\frac{g_{\pi \ssst{NN}}}{2 \mN} \right)^2 (\bm{\tau}_1 \times \bm{\tau}_2)^z
(\bm{\sigma}_1\cdot\nabla_{1})\ (\bm{\sigma}_2\cdot\nabla_{2})\
(\nabla_{1} - \nabla_{2})\ f_\pi(r_{x1})\ f_\pi(r_{x2}).
\end{eqnarray}
\bibliographystyle{apsrev}
\bibliography{AM,MEC,QEPV}

\end{document}